# Using the Makerspace to Create Educational Open-source Software for Electrical Circuits: A Learning Experience

Dana Conard[1], Blake Vollmer[2], Corbin Shatto[3], Hannah Bowman[4], and Sara Kassis[5]

[1]Dana Conard; Dept. of Comp. Sci., Sonoma State University; e-mail: conard@sonoma.edu
[2]Blake Vollmer; Dept. of Comp. Sci., Sonoma State University; e-mail: vollmerb@sonoma.edu
[3]Corbin Shatto; Dept. of Eng. Sci., Sonoma State University; e-mail: shatto@sonoma.edu
[4]Hannah Bowman; Dept. of Anthrop., Sonoma State University; e-mail: bowmanh@sonoma.edu
[5]Sara Kassis; Dept. of Eng. Sci., Sonoma State University; e-mail: sara.kassis@sonoma.edu

**Introduction**

There are advantages in using virtual reality to overcome challenges in education [1], [2], [3]. This is especially useful in addressing methods to increase student success such as use in courses where a high number of students obtain a repeatable grade (D, F, or Withdraw) or in bottleneck courses where limitations such as class size or lack of resources impose enrollment restrictions. Utilizing such a modality would indeed align with the California State University's focus to increase student success rates and to eliminate opportunity and achievement gaps [4].

Virtual learning environments are a useful modality for engaging students in the classroom by affording them a sense of presence and immersion. This, in turn, enables students to learn tasks that are impossible or not effective in a typical learning environment [1]. Additionally, students are able to retain the information better when a virtual learning environment is used to complement the traditional learning environment [5].

As for the electrical engineering discipline, there is evidence to show that students have difficulty in understanding fundamental behavior in electrical circuits [6]. Additionally, it is shown that performance in the lower division coursework on a student's university education when majoring in electrical engineering is a strong indicator for whether a student graduates [7]. It is, therefore, the motivation of this project to create an open-source virtual and augmented reality electrical circuit application to be used in lower division engineering courses to teach students about electricity fundamentals. Upon completion, this project will be used in future research projects to measure the efficacy of such a modality on a student's education.

**Virtual and Augmented Reality Development Using the Makerspace**

A. *Virtual Immersive Teaching and Learning Laboratory*

A newly formed Virtual Immersive Teaching and Learning (VITaL)* Laboratory (see Fig. 1) was a recent addition to the current Makerspace at Sonoma State University, a liberal arts institution, where students from all disciplines are able to utilize virtual reality head-mounted displays (HMD), augmented reality devices, 360-degree cameras, and VR-capable computers as a part of their academic experience.

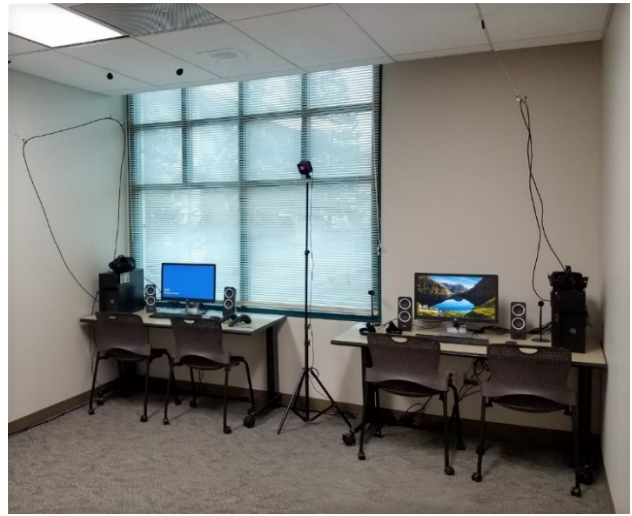

*Fig. 1 Part of the Virtual Immersive Teaching and Learning Laboratory in the Makerspace*

Students have been using this space to learn more about their subject material in virtual or augmented reality in addition to their classroom instruction by using already-created virtual and augmented reality applications with instructions written by their faculty member to compliment course content and enhance their learning experience. This space is considered a project space according to [8].

This same laboratory is also used as a community space [8] by Sonoma State University's VR Club, a student-led group open to all majors which meets to experience virtual and augmented reality.

The software that is readily available for use on virtual and augmented reality devices does not necessarily apply to all disciplines and does not necessarily have a pedagogical or accessibility focus. Considering this lack of appropriate educational applications for the current virtual and augmented reality devices, a team of interdisciplinary students was formed to create this software as a summer project. It is currently in development and is anticipated to be finished by the end of summer. This project not only serves the purpose of creating this much needed software, but also offers these students a valuable experience in gaining skill sets for their future careers.

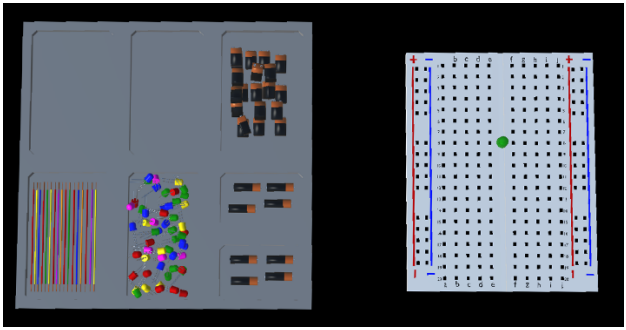

*Fig. 2 Electric circuit application interface including a toolbox and breadboard*

This project will make use of the VITaL Laboratory within the University's Makerspace in using the hardware to create and test this application in a project-based effort.

### B. *Hardware and Software Used*

The VITaL Laboratory contains virtual reality devices HTC VIVE and Oculus Rift, mixed reality devices (considered a subset of augmented reality) Microsoft HoloLens, and VR-capable computers. The electric circuit software is currently being created to utilize the HoloLens and future plans include modifying the software to run on other platforms. The HoloLens is chosen specifically because of its possibility to see physical reality with the computer generated images, giving students the choice to overlay the electric circuits application on top of a real circuit to further visualize current flow and magnetic field. The students will use the hardware in the VITaL Laboratory to build and test the software they are creating. The student developers are using Unity and Visual Studio for the development and Blender for the 3D modeling.

### Student Development Team

The interdisciplinary student development team positions are modeled much like a professional software team where each person contributes according to their expertise**:** Two computer science students are creating the 3D models and writing the software code; one electrical engineering student is the subject matter expert and consults with the rest of the team members on the fundamental properties of the electric circuit; one anthropology student is the user experience designer and will offer expertise on accessibility in addition to consulting on the user interface; lastly, a faculty member is the project lead. The students are all junior and senior level standing in their discipline.

### Electric Circuits Application

The electric circuits application will allow students to create a virtual electric circuit, much like what they physically build in class, except they will now be able to visualize the electron flow and associated magnetic field. These are two concepts that are difficult for students to visualize and yet constitute an essential part in understanding electric circuits.

The project consists of an interface where students can use an electronic component tool box to build a virtual electric circuit on a virtual breadboard (see Fig. 2). In the toolbox, students can find resistors, LEDs, diodes, capacitors, transistors, DC batteries, an AC voltage source, and wires. Students will be able to interact with the toolbox and move components onto a virtual breadboard to virtually build a circuit. Once the circuit is in a closed loop, virtual electrons will appear and move accordingly, showing the student the behavior of electricity fundamentals. The electron flow rate will be dependent on the values a student inputs for the power, the type of power source, and the quantity and positioning of the electronic components. Additionally, a corresponding magnetic field will appear once the current is flowing through the wire. The strength of the magnetic field will reflect on the input values for the electric components and the power sources.

Fig. 2 shows the latest interface that has been created thus far. This includes a toolbox with wires, LEDs, and various sizes of batteries. To the right of this is a virtual breadboard onto which students place a component from the toolbox. This a work in progress and a demo version showing the electric current and magnetic field will be available at the completion of the summer.

### In Summary

There is a lack of appropriate virtual and augmented reality applications available for use as a learning modality in education. Therefore, it is necessary to establish a multi-disciplinary, in-house team of students, led by faculty, to develop this much-needed educational software. This opportunity also offers students a unique learning experience in creating the software being used.

The tools being used to create this software application are a part of the Virtual Immersive Teaching and Learning Lab which is housed in the Sonoma State University Makerspace. This project is focused on building a necessary application for introductory electrical engineering classes. It is currently in development and is projected to be completed towards the end of summer break.

Using this interactive electric circuit in the classroom and observing its efficacy on student success will be the subject of a future research project.

### Acknowledgements

The authors would like to thank the various leadership at Sonoma State University whom have helped make this possible, including those at the Office of the Provost, Information Technology, Library, School of Science and Technology, and the Faculty Center.

*The name Virtual Immersive Teaching and Learning is attributed in partnership with San Diego State University.

### References

[1] M. L. B. Dalgarno, "What are the learning affordances of 3-D virtual environments?," *British Journal of Educational Technology,* vol. 41, no. 1, p. 10–32, 2010.

[2] J. Psotka, "Educational Games and Virtual Reality as Disruptive Technologies," *Journal of Educational Technology & Society,* vol. 16, no. 2, pp. 69 - 80, 2013.

[3] K. P. A. N. R. B. S. J. &. G. K. J. Bailenson, "The Effect of Interactivity on Learning Physical Actions in Virtual Reality," *Media Psychology,* vol. 11, no. 3, pp. 354-376, 2008.

[4] California State University, 2018. [Online]. Available: https://www2.calstate.edu/csu-system/why-the-csu-matters/graduation-initiative-2025.

[5] S. C. J. W. J. Crosier, "Experimental Comparison of Virtual Reality with Traditional Teaching Methods for Teaching Radioactivity," *Education and Information Technologies,* vol. 5, no. 4, p. 329 – 343, 2000.

[6] B. E. a. U. G. R. Cohen, "Potential difference and current in simple electric circuits: A study of students' concepts," *American Journal of Physics,* vol. 51, no. 407, 1983.

[7] B. S. C. M. A. S. C. S. E. O. S. Sajjadi, "Finding Bottlenecks: Predicting Student Attrition with Unsupervised Classifier," *IEEE, IntelliSys,* 2017.

[8] M. Culpepper, "Types of Academic Makerspaces, Their Import to the Education Mission, and The Characteristics of Their Culture and Community," *Proceedings of the 1st International Symposium on Academic Makerspaces,* pp. 10 - 13, 2016.